\documentclass[letter]{aa} 

\usepackage{natbib}
\usepackage{graphicx}
\usepackage{txfonts}
\begin{document}

\title{Optical circular polarization of blazar S4 0954+65 during high linear polarized states}


\author{I. Liodakis\inst{1,2}\thanks{yannis.liodakis@gmail.com}, E. Shablovinskaya\inst{3,4}, D. Blinov\inst{5,6},  S. S. Savchenko\inst{3,7,8},  E. Malygin\inst{3}, S. Kotov\inst{3}, S. Kiehlmann\inst{5,6}, A. C. S. Readhead\inst{9}, S. B. Potter\inst{10,11}, F. M. Rieger\inst{12,13}, T. S. Grishina\inst{7}, V. A. Hagen-Thorn\inst{7}, E. N. Kopatskaya\inst{7}, E. G. Larionova\inst{7}, D. A. Morozova\inst{7}, I. S. Troitskiy\inst{7}, Y. V. Troitskaya\inst{7}, A. A. Vasilyev\inst{7}, A. V. Zhovtan\inst{14}, \and G. A. Borman \inst{14}}
          
\institute{$^1$Finnish Centre for Astronomy with ESO, 20014 University of Turku, Finland \\
$^2$NASA Marshall Space Flight Center, Huntsville, AL 35812, USA\\
$^3$Special astrophysical observatory of Russian Academy of Sciences, Nizhnĳ Arkhyz, Karachai-Cherkessian Republic, 369167, Russia\\
$^4$Instituto de Estudios Astrof\'isicos, Facultad de Ingenier\'ia y Ciencias, Universidad Diego Portales, Santiago, Regi\'on Metropolitana, 8370191 Chile\\
$^5$Institute of Astrophysics, Foundation for Research and Technology-Hellas, GR-70013 Heraklion, Greece\\
$^6$Department of Physics, University of Crete, 70013, Heraklion, Greece\\
$^7$Saint Petersburg State University, 7/9 Universitetskaya nab., St. Petersburg, 199034 Russia\\
$^8$Pulkovo Observatory, St.Petersburg, 196140, Russia\\
$^{9}$Owens Valley Radio Observatory, California Institute of Technology, Pasadena, CA 91125, USA\\
$^{10}$South African Astronomical Observatory, PO Box 9, Observatory, 7935, Cape Town, South Africa\\
$^{11}$Department of Physics, University of Johannesburg, PO Box 524, Auckland Park 2006, South Africa\\
$^{12}$Institute for Theoretical Physics (ITP), University of Heidelberg, Philosophenweg 12, 69120 Heidelberg, Germany\\
$^{13}$Max Planck Institute for Plasma Physics, Boltzmannstraße 2,
85748 Garching, Germany\\
$^{14}$Crimean Astrophysical Observatory RAS, P/O Nauchny, 298409, Crimea\\
}

\titlerunning{Optical Circular polarization of S4~0954+65}
\authorrunning{Liodakis et al.,}

 
  \abstract{Optical circular polarization observations can directly test the particle composition in black holes jets. Here we report on the first observations of the BL Lac type object S4 0954+65 in high linear polarized states. While no circular polarization was detected, we were able to place upper limits of $<0.5\%$ at the 99.7\% confidence. Using a simple model and our novel optical circular polarization observations we can constrain the allowed parameter space for the magnetic field strength and composition of the emitting particles. Our results favor models that require magnetic field strengths of  only a few Gauss and models where the jet composition is dominated by electron-positron pairs. We discuss our findings in the context of typical magnetic field strength requirements for blazar emission models.}  


   \maketitle
%

\section{Introduction} \label{sec:intro}
The origin of the high-energy (keV - TeV) emission in blazar jets is a highly debated open question since their discovery \citep{Blandford2019,Hovatta2019}. In the past few years that debate has intensified since the potential association of TXS~0506+06 with a high-energy neutrino \citep{icecube2018}. The standard approach to the problem has been modeling of the (quasi-)simultaneous spectral energy distribution (SED) of flaring blazars (e.g., \citealp{Raiteri2017-II,magic2022,magic2023}). However,  the incomplete wavelength coverage, the complexity of even the simplest models, and the uncertainty introduced by relativistic effects and degenerate parameters leads to multiple - often drastically different - models to provide equally good fits to the data. 

X-ray polarization has recently provided a new path forward with the launch of the Imaging X-ray Polarimetry Explorer (IXPE, \citealp{Weisskopf2022}). The first observations of IXPE clearly demonstrated two things. First, the most commonly used single-zone models to describe the SEDs are not consistent with the observed behavior \citep{Liodakis2022,DiGesu2023,Middei2023-II}. Second, the X-ray polarization observations of low synchrotron peaked sources (LSP, i.e., where the X-ray emission is dominated by the high-energy emission component) favor a scenario where inverse-Compton scattering from relativistic electrons is likely responsible for at least the keV emission in LSP-blazars \citep{Middei2023,Peirson2023}. However, LSPs are typical at or below the detection capabilities of IXPE, hence it is likely that future missions like e-XTP would be required to solve this puzzle \citep{Peirson2022}.

All of the above demonstrate that new and complementary approaches are necessary. Optical circular polarization (OCP) offers such a complementary pathway, however, to this day it remains unexplored due to the low number of optical polarimeters available with CP capabilities. There are only a handful of OCP measurements for blazars, and quasars in general \citep{Valtaoja1993,Wagner2001,Hutsemekers2010,Liodakis2022-II}, all of which are $\leq1\%$. While the vast majority are upper limits, there are three sources (3C~279, \citealp{Wagner2001}; PKS~1256-229, PKS~2155-152, \citealp{Hutsemekers2010}) with a $>3\sigma$ detection. However, it is still unclear if the reported OCP values are due to instrumental effects \citep{Bagnulo2011}, and if the origin of OCP is intrinsic to the sources (for discussion on alternative models see \citealp{Wagner2001,Rieger2005}).

Here we report on the first OCP observations of the LSP\footnote{S4~0954+65 has a synchrotron peak frequency of $\nu_{syn}=1.1\times10^{13}$~Hz \citep{Ajello2020}} BL Lac-type Object S4~0954+65. S4~0954+65 sits at $\rm RA=09h58m47.24s$, Dec=+65$\rm^o$33$^\prime$54.81$^{\prime\prime}$ and z=0.367. It is one of the few LSP blazars that has been detected in TeV $\gamma$-rays \citep{MAGIC2018}, often showing a very high degree ($>30\%$) of optical linear polarization (OLP, e.g., \citealp{Blinov2021,Raiteri2021}). In section \ref{sec:data_model} we discuss the OCP observations and attempt to constrain the magnetic field strength and particle composition of the jet. In section \ref{sec:discussion} we discuss our findings.

\section{Optical circular polarization observations and modeling} \label{sec:data_model}

OCP observations were performed with the SCORPIO-2 focal reducer \citep{scorpio2} with thick CCD E2V 261-84 \citep{afanasieva23} at the 6m BTA telescope of Special astrophysical observatory in highly-polarized OLP states found by the St. Petersburg State University's monitoring program \footnote{\url{https://vo.astro.spbu.ru/program}}. 
In OCP mode, we observed S4 0954+65 twice, in June 2022 and in May 2023. In both cases, we used a z-SDSS broad-band filter ($\Delta \lambda \sim$830-1000 nm). The double Wollaston prism was used as a polarization analyzer with the $\lambda/4$ phase plate used in the fixed positions of 0$^\circ$ and 90$^\circ$. The observations were conducted under non-clear weather conditions with the cirruses and the seeing of $\sim$2$''$ and $\sim$1.2$''$ in the first and in the second epoch, respectively. Nevertheless, the signal-to-noise ratio obtained from each frame with 20-sec exposure was S/N $\sim$~500 in 2022 and 400 in 2023. The total integration time is 1.06 hours and 1.39 hours, respectively. 
The data was reduced following the standard steps of bias and flat field subtraction, as well as the correction of the polarization channels transmission performed using two stars in 3$'$ FoV around the source. The processing of the SCORPIO-2 polarimetric data is described in detail in \cite{afanasiev12}. Table \ref{tab:Table1} summarizes the OCP and OLP observations.

\begin{table*}
\setlength{\tabcolsep}{11pt}
\centering
  \caption{Summary of observations for S4~0954+65. The columns are (1) Julian Date, (2) Gregorian date, (3) linear polarization degree (\%), (4) polarization angle (degrees), and (5) upper limit on the circular polarization degree (\%) at the 99.7\% confidence interval. The linear polarization observations are in the R-band and the circular polarization observations in the z-SDSS band.}
  \label{tab:Table1}
\begin{tabular}{@{}ccccc@{}}
 \hline
 JD  & Date & $\Pi_{l}$  & PA & $\Pi_{c}$  \\
  \hline
2459738.3737 & 2022-06-07 & 30.9$\pm$0.5 & 83$\pm$0.5 & $<$0.184 \\
2460084.4184 & 2023-05-19 & 14.6$\pm$0.7 & 135$\pm$1 & $<$0.453 \\
\hline
\end{tabular}
\end{table*}

Following \cite{Liodakis2022-II} we constrain the magnetic field strength and positron fraction using, 
\begin{equation}
B \approx 2\times 10^7 \left( \frac{\nu_{\rm obs}}{10^{15}~Hz}\right)   \left( \frac{0.71}{\Pi_l}\right)^{2}  \left( \frac{1}{\Gamma^3(1-2f)^2}\right)  \,     \Pi_c^2,
\label{Eq:bfield}
\end{equation}
where, $\Pi_l$ and $\Pi_c$ are the OLP and OCP respectively,  $f$ is the fraction of positrons, $\nu_{\rm obs}$ is the observing frequency set to the central wavelength of the z-SDSS band (3.28$\times10^{14}$~Hz), and $B$ the intrinsic magnetic field strength in Gauss \citep{Rieger2005}. $f$ is defined as the fraction of positrons to the total number of leptons, i.e., $f=N_{e^+}/(N_{e^+} + N_{e^-})$. Hence, $f=0$ is for a pure proton-electron plasma (a.k.a. normal plasma) and $f=0.5$ is for a pure electron-positron plasma (a.k.a. pair plasma). The model assumes a power law distribution of emitting particles, a perpendicular to the jet axis intrinsic magnetic field, and the jet viewed at the critical angle ($\theta_{obs}\approx1/\Gamma$), which implies that the Lorentz factor ($\Gamma$) is approximately equal to the Doppler factor ($\Gamma\approx\delta$).  We estimate the Doppler factor of the source following \cite{Liodakis2018} and using the most recent data release from the Owens Valley Radio Observatory (up to August 2023, \citealp{Richards2011}). In brief, we use the Bayesian Hierarchical models implemented in {\it Magnetron}\footnote{\url{https://github.com/dhuppenkothen/magnetron2/tree/blazars}} \citep{Huppenkothen2015} to model the radio light curve at 15~GHz as a series of flares superimposed on stochastic backgrounds. We use  {\it Magnetron}'s results to compute the maximum observed brightness temperature distribution which we compare to the intrinsic maximum brightness temperature of $T_{int,max}= 2.78\times10^{11}$~K \citep{Liodakis2018}. We find a Doppler factor of $\delta=11.6^{+11.1}_{-5.5}$ consistent with previous estimates in the literature \citep{Liodakis2017,Liodakis2021}. Note that the uncertainties on $\delta$ are not statistical, but rather express the range of possible values over the $\sim$15 years of radio observations. In addition, the imperfectly ordered magnetic field will reduce OCP and OLP similarly from the maximum polarization (\citealp{Jones1977}, hence the $[0.71/\Pi_l]$ term). The OLP observations are in the R-band, however, the polarization degree of S4~0954+65 is fairly achromatic (e.g., \citealp{Raiteri2021}) with any differences far below the observational uncertainties, which we have taken into account in our simulations below.

\begin{figure*}
\centering
\includegraphics[width=0.45\textwidth]{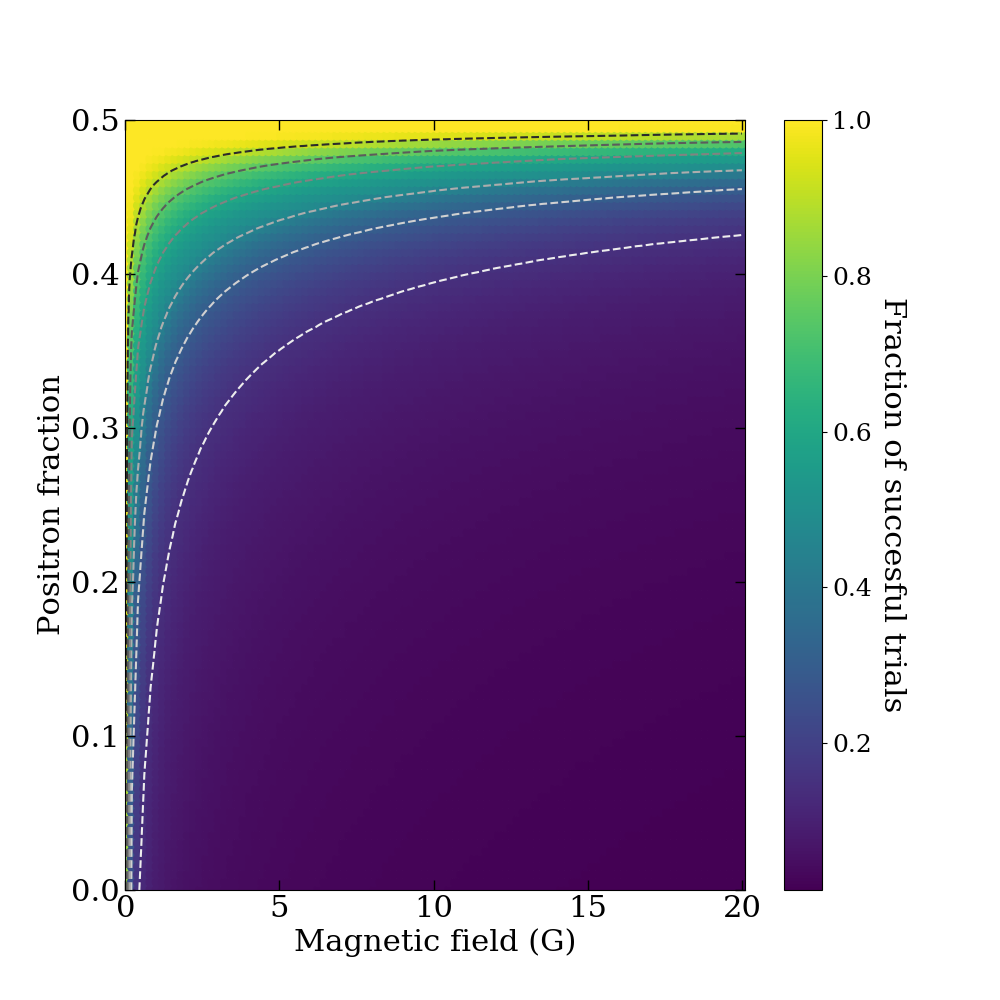}
\includegraphics[width=0.45\textwidth]{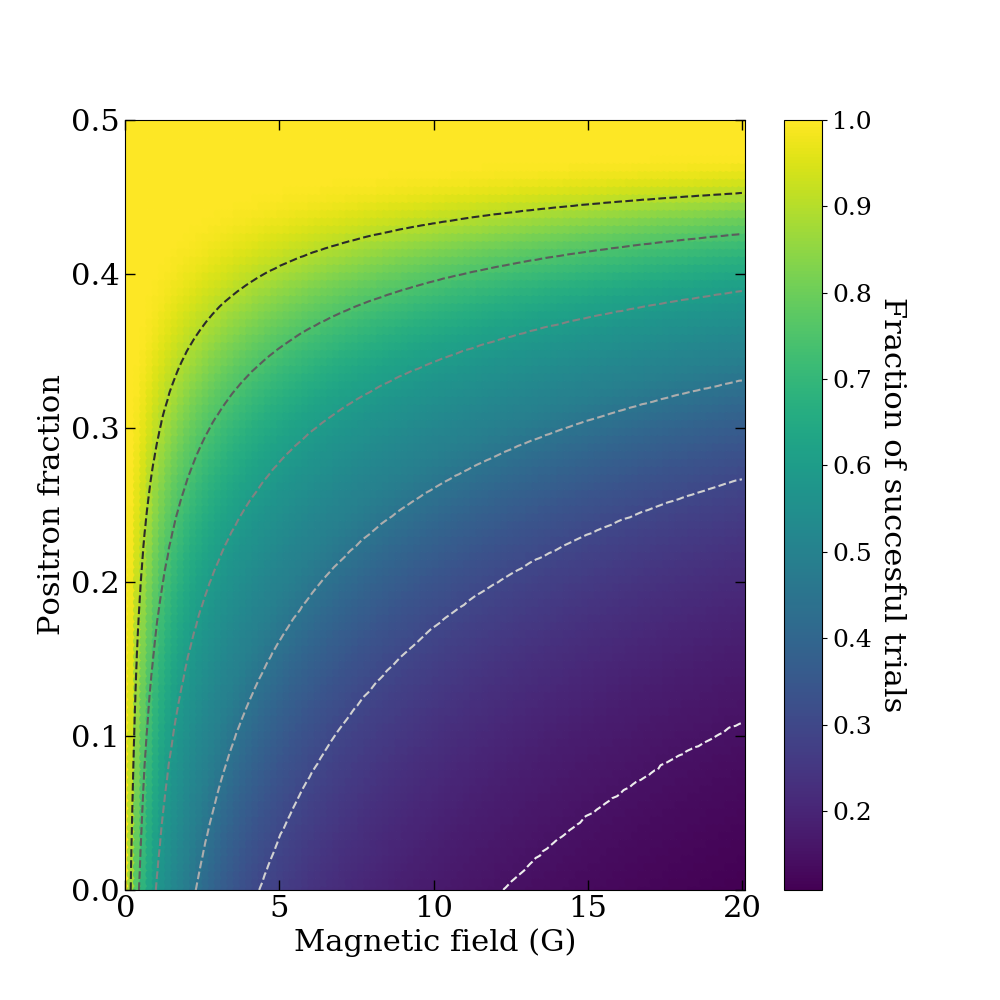}
\caption{Constraints on the magnetic field strength and positron fraction for both observations of S4 0954+65. The left panel is for JD~2459738.3737 and the right panel for JD~2460084.4184. The colorbar shows the fraction of successful simulations for each ($B,~f$) pair. The contours have been added to guide the reader's eye.}
\label{plt:constraints}
\end{figure*}

Given the above model, we explore a parameter space for $f$=[0,0.5] and for $B$=[0.1,20] Gauss. For each pair of ($B,~f$) we draw a random value for the Doppler factor and the OLP given the uncertainties. We marginalize over all potential values by repeating the process 20,000 times. For each simulation, we evaluate if the produced OCP is below the observed 3$\sigma$ upper limit. Figure \ref{plt:constraints} shows the results of the simulations. It is evident for both observations that models with magnetic field strengths $B>10$ Gauss and low positron fraction ($f<0.2$) are disfavored. It is also evident that the observation on JD~2459738.3737 provides more stringent constraints on [$B,~f$].

\section{Discussion} \label{sec:discussion}

We presented the first optical circular polarization observations of S4 0954+65 in high linear polarized states. Our observations provide strict upper limits below $<0.5\%$ at the 3$\sigma$ level. The non-detection of the OCP at such low level is consistent with our previous observations on 3C~279 and PKS~1510-089 \citep{Liodakis2022-II}, as well as other past blazar observations \citep{Valtaoja1993,Wagner2001,Hutsemekers2010}. We re-estimated the Doppler factor for the source using $\sim$15 years of radio observations at 15~GHz from OVRO, which we use to constrain its magnetic field strength and jet composition. Throughout our modeling we have assumed that the optical and radio emission regions have the same Doppler factor, and all the potential OCP signal is intrinsic to the source. If there is contribution from the interstellar dust to the OCP, then our limits will decrease, tightening the constraints on ($B,~f$). Similarly, if the optical emission region has a higher Doppler factor, as typically assumed to resolve the Doppler-crisis in high-synchrotron peak blazars\footnote{i.e., blazars with $\nu_{syn}>10^{15}$~Hz} \cite[e.g.,][]{Georganopoulos2003}, our constraints would tighten further. Our model considers a scenario where the intrinsic magnetic field is perpendicular to the jet axis as e.g., expected in shock-in jet scenarios \citep{Marscher1985}. In such a case, the intrinsic CP is maximized. Our scenario would suggest the polarization angle to be roughly parallel to the jet, which has been found to be true in individual source studies and small sample analyses \cite[e.g.,][]{Hovatta2016}. Recent X-ray polarization results on high-peaked blazars (analogous to optical polarization for low-peaked blazars) have also found the X-ray polarization angle to be about the jet axis \citep{Liodakis2022,DiGesu2022}. However, the position angle of the jet can vary over time \cite[e.g.,][]{Lico2020}, the optical polarization angle is often highly variable \citep{Kiehlmann2021}, and to our knowledge, there is no population study on the connection between the magnetic field orientation and the jet direction. The radio polarization angle in S4~0954+65 has been found to be parallel to the jet axis \citep{osullivan2008,Hodge2018}. The position angle of the jet is $166.3\pm2.1^\circ$ \citep{Weaver2022}, however, it varies from $129^\circ$ to $174^\circ$ \cite[e.g.,][]{MAGIC2018}. Near in time to our observations, the innermost jet direction changes from $168\pm3^\circ$ (2022-06-05) to $129\pm3^\circ$ (2023-05-21, S. Jorstad -- private communication).  Based on our OLP angle measurements (Table \ref{tab:Table1}), the magnetic field was roughly parallel to the jet during the first observation and perpendicular during the second. The OLP angle traces the integrated magnetic field direction along the line of sight and the inner jet direction is estimated using the brightest knot that is well-resolved from the radio core (i.e., the brightest feature in the radio image). Hence, our estimates might not be representative of the true direction of the magnetic field with respect to the jet axis. While our quantitative estimates might be affected, given that our observations encompass different conditions, it is unlikely that our conclusions will be altered.

We find that large magnetic field values ($>10$~Gauss) and predominately proton-electron jets overproduce the observed level of OCP. This is important in the context of leptonic versus hadronic models, since the latter typically require much larger magnetic field strengths ($>30$~Gauss, e.g., \citealp{Cerruti2020}). \cite{Liodakis2020-II} estimated the minimum magnetic field strength required for proton synchrotron models to produce the observed SED for 145 blazars. While S4~0954+65 was not included in their sample, we can use the distribution of the population. For all subclasses of blazars we find a median of 238~Gauss and the lowest estimate to be 2.85~Gauss. Restricting the sample to the LSP sources (such as S4~0954+65) then the median and lowest value are 243.2~Gauss and 5.34~Gauss respectively. For high values close to the median, we can exclude the pure electron-proton models at $>4\sigma$ significance. In fact, for any value above 70~Gauss we can exclude the pure electron-proton models at $>3\sigma$ significance. At 5.34~Gauss, as the minimum found for LSPs, only 3\% of the simulations can produce consistent OCP values. For 2.85~Gauss, 4.4\% of the simulations are consistent with the observations.

Estimates on the magnetic field strength of S4 0954+65 through leptonic SED modeling \citep{Tanaka2016,MAGIC2018} suggest values at 1~Gauss or lower. In that case, $>9\%$ of the simulations can produce consistent OCP values. Therefore, we can not exclude the presence of a significant fraction of relativistic protons. However, at such low magnetic field strengths, protons are no longer efficient in producing radiation \citep{Sikora2009}. We note that these B-field estimates are based on single-zone models which are strongly disfavored by the recent X-ray polarization observations \cite[e.g.,][]{Liodakis2022}, but are in fact typically found independently from radio observations \cite[e.g.,][]{Pushkarev2012}. If this is the case, then it would suggest that proton emission, and the potential high-energy neutrino emission, is rare or takes place under certain conditions. For example, compression from a shock that would amplify the magnetic field strength, shock-shock interactions or rotations of the polarization angle (e.g., \citealp{Liodakis2020,Novikova2023}). This is in-line with the small fraction of observed orphan $\gamma$-ray flares \citep{Liodakis2019,deJaeger2023}, the predictions for neutrino emission from X-ray/$\gamma$-ray flares \citep{Oikonomou2019,Stathopoulos2022}, and could potentially explain why the neutrino-blazar connection remains uncertain even after significant effort \cite[e.g.,][]{Plavin2020,Hovatta2021,Kovalev2023}.

\begin{acknowledgements}
We thank the anonymous referee for comments that helped improve this work. We also thank Svetlana Jorstad for providing the estimates of the inner jet position angles at the time of our observations through the BEAM-ME program. I. Liodakis was supported by the NASA Postdoctoral Program at the Marshall Space Flight Center, administered by Oak Ridge Associated Universities under contract with NASA. This research has made use of data from the OVRO 40-m monitoring program \citep{Richards2011}, supported by private funding from the California Institute of Technology and the Max Planck Institute for Radio Astronomy, and by NASA grants NNX08AW31G, NNX11A043G, and NNX14AQ89G and NSF grants AST-0808050 and AST- 1109911. S. Kiehlmann acknowledges support from the European Research Council (ERC) under the European Unions Horizon 2020 research and innovation programme under grant agreement No.~771282.  E.S. Shablovinskaya acknowledges support from ANID BASAL project FB210003 and Gemini ANID ASTRO21-0003. Observations with the SAO RAS telescopes are supported by the Ministry of Science and Higher Education of the Russian Federation. The renovation of telescope equipment is currently provided within the national project "Science and Universities". E.S. Shablovinskaya, E.A. Malygin and S.S. Kotov obtained observational data on the unique scientific facility "Big Telescope Alt-azimuthal" of SAO RAS as well as made data processing with the financial support of grant No075-15-2022-262 (13.MNPMU.21.0003) of the Ministry of Science and Higher Education of the Russian Federation. This study makes use of VLBA data from the VLBA-BU Blazar Monitoring Program (BEAM-ME and VLBA-BU-BLAZAR; \url{http://www.bu.edu/blazars/BEAM-ME.html}), funded by NASA through the Fermi Guest Investigator Program. The VLBA is an instrument of the National Radio Astronomy Observatory. The National Radio Astronomy Observatory is a facility of the National Science Foundation operated by Associated Universities, Inc.
\end{acknowledgements}

\bibliographystyle{aa} 

\begin{thebibliography}{55}
\expandafter\ifx\csname natexlab\endcsname\relax\def\natexlab#1{#1}\fi

\bibitem[{{Acciari} {et~al.}(2022){Acciari}, {Aniello}, {Ansoldi}, {Antonelli},
  {Arbet Engels}, {Artero}, {Asano}, {Baack}, {Babi{\'c}}, {Baquero}, {Barres
  de Almeida}, {Barrio}, {Batkovi{\'c}}, {Becerra Gonz{\'a}lez}, {Bednarek},
  {Bernardini}, {Bernardos}, {Berti}, {Besenrieder}, {Bhattacharyya},
  {Bigongiari}, {Biland}, {Blanch}, {B{\"o}kenkamp}, {Bonnoli},
  {Bo{\v{s}}njak}, {Busetto}, {Carosi}, {Ceribella}, {Cerruti}, {Chai},
  {Chilingarian}, {Cikota}, {Colombo}, {Contreras}, {Cortina}, {Covino},
  {D'Amico}, {D'Elia}, {Vela}, {Dazzi}, {De Angelis}, {De Lotto}, {Del Popolo},
  {Delfino}, {Delgado}, {Mendez}, {Depaoli}, {Di Pierro}, {Di Venere}, {Do
  Souto Espi{\~n}eira}, {Dominis Prester}, {Donini}, {Dorner}, {Doro},
  {Elsaesser}, {Fallah Ramazani}, {Fari{\~n}a}, {Fattorini}, {Font}, {Fruck},
  {Fukami}, {Fukazawa}, {Garc{\'\i}a L{\'o}pez}, {Garczarczyk}, {Gasparyan},
  {Gaug}, {Giglietto}, {Giordano}, {Gliwny}, {Godinovi{\'c}}, {Green}, {Green},
  {Hadasch}, {Hahn}, {Hassan}, {Heckmann}, {Herrera}, {Hoang}, {Hrupec},
  {H{\"u}tten}, {Inada}, {Iotov}, {Ishio}, {Iwamura}, {Jim{\'e}nez
  Mart{\'\i}nez}, {Jormanainen}, {Jouvin}, {Kerszberg}, {Kobayashi}, {Kubo},
  {Kushida}, {Lamastra}, {Lelas}, {Leone}, {Lindfors}, {Linhoff}, {Lombardi},
  {Longo}, {L{\'o}pez-Coto}, {L{\'o}pez-Moya}, {L{\'o}pez-Oramas}, {Loporchio},
  {Machado de Oliveira Fraga}, {Maggio}, {Majumdar}, {Makariev}, {Mallamaci},
  {Maneva}, {Manganaro}, {Mannheim}, {Mariotti}, {Mart{\'\i}nez}, {Mas
  Aguilar}, {Mazin}, {Menchiari}, {Mender}, {Mi{\'c}anovi{\'c}}, {Miceli},
  {Miener}, {Miranda}, {Mirzoyan}, {Molina}, {Moralejo}, {Morcuende}, {Moreno},
  {Moretti}, {Nakamori}, {Nava}, {Neustroev}, {Nievas Rosillo}, {Nigro},
  {Nilsson}, {Nishijima}, {Noda}, {Nozaki}, {Ohtani}, {Oka}, {Otero-Santos},
  {Paiano}, {Palatiello}, {Paneque}, {Paoletti}, {Paredes}, {Pavleti{\'c}},
  {Pe{\~n}il}, {Persic}, {Pihet}, {Prada Moroni}, {Prandini}, {Priyadarshi},
  {Puljak}, {Rhode}, {Rib{\'o}}, {Rico}, {Righi}, {Rugliancich}, {Sahakyan},
  {Saito}, {Sakurai}, {Satalecka}, {Saturni}, {Schleicher}, {Schmidt},
  {Schmuckermaier}, {Schweizer}, {Sitarek}, {{\v{S}}nidari{\'c}}, {Sobczynska},
  {Spolon}, {Stamerra}, {Stri{\v{s}}kovi{\'c}}, {Strom}, {Strzys}, {Suda},
  {Suri{\'c}}, {Takahashi}, {Takeishi}, {Tavecchio}, {Temnikov}, {Terzi{\'c}},
  {Teshima}, {Tosti}, {Truzzi}, {Tutone}, {Ubach}, {van Scherpenberg}, {Vanzo},
  {Vazquez Acosta}, {Ventura}, {Verguilov}, {Viale}, {Vigorito}, {Vitale},
  {Vovk}, {Will}, {Wunderlich}, {Yamamoto}, {Zari{\'c}}, {Hodges}, {Hovatta},
  {Kiehlmann}, {Liodakis}, {Max-Moerbeck}, {Pearson}, {Readhead}, {Reeves},
  {L{\"a}hteenm{\"a}ki}, {Tornikoski}, {Tammi}, {D'Ammando}, \&
  {Marchini}}]{magic2022}
{Acciari}, V.~A., {Aniello}, T., {Ansoldi}, S., {et~al.} 2022, \apj, 927, 197

\bibitem[{{Afanasiev} \& {Amirkhanyan}(2012)}]{afanasiev12}
{Afanasiev}, V.~L. \& {Amirkhanyan}, V.~R. 2012, Astrophysical Bulletin, 67,
  438

\bibitem[{{Afanasiev} \& {Moiseev}(2011)}]{scorpio2}
{Afanasiev}, V.~L. \& {Moiseev}, A.~V. 2011, Baltic Astronomy, 20, 363

\bibitem[{{Afanasieva} {et~al.}(2023){Afanasieva}, {Murzin}, {Ardilanov},
  {Ivaschenko}, {Pritychenko}, {Moiseev}, {Shablovinskaya}, \&
  {Malygin}}]{afanasieva23}
{Afanasieva}, I., {Murzin}, V., {Ardilanov}, V., {et~al.} 2023, Photonics for
  Solar Energy Systems IX, 10, 774

\bibitem[{{Ajello} {et~al.}(2020){Ajello}, {Angioni}, {Axelsson}, {Ballet},
  {Barbiellini}, {Bastieri}, {Becerra Gonzalez}, {Bellazzini}, {Bissaldi},
  {Bloom}, {Bonino}, {Bottacini}, {Bruel}, {Buson}, {Cafardo}, {Cameron},
  {Cavazzuti}, {Chen}, {Cheung}, {Ciprini}, {Costantin}, {Cutini}, {D'Ammando},
  {de la Torre Luque}, {de Menezes}, {de Palma}, {Desai}, {Di Lalla}, {Di
  Venere}, {Dom{\'\i}nguez}, {Dirirsa}, {Ferrara}, {Finke}, {Franckowiak},
  {Fukazawa}, {Funk}, {Fusco}, {Gargano}, {Garrappa}, {Gasparrini},
  {Giglietto}, {Giordano}, {Giroletti}, {Green}, {Grenier}, {Guiriec},
  {Harita}, {Hays}, {Horan}, {Itoh}, {J{\'o}hannesson}, {Kovac'evic'},
  {Krauss}, {Kreter}, {Kuss}, {Larsson}, {Leto}, {Li}, {Liodakis}, {Longo},
  {Loparco}, {Lott}, {Lovellette}, {Lubrano}, {Madejski}, {Maldera},
  {Manfreda}, {Mart{\'\i}-Devesa}, {Massaro}, {Mazziotta}, {Mereu}, {Meyer},
  {Migliori}, {Mirabal}, {Mizuno}, {Monzani}, {Morselli}, {Moskalenko},
  {Negro}, {Nemmen}, {Nuss}, {Ojha}, {Ojha}, {Omodei}, {Orienti}, {Orlando},
  {Ormes}, {Paliya}, {Pei}, {Pe{\~n}a-Herazo}, {Persic}, {Pesce-Rollins},
  {Petrov}, {Piron}, {Poon}, {Principe}, {Rain{\`o}}, {Rando}, {Rani},
  {Razzano}, {Razzaque}, {Reimer}, {Reimer}, {Schinzel}, {Serini}, {Sgr{\`o}},
  {Siskind}, {Spandre}, {Spinelli}, {Suson}, {Tachibana}, {Thompson}, {Torres},
  {Torresi}, {Troja}, {Valverde}, {van Zyl}, \& {Yassine}}]{Ajello2020}
{Ajello}, M., {Angioni}, R., {Axelsson}, M., {et~al.} 2020, \apj, 892, 105

\bibitem[{{Bagnulo} {et~al.}(2011){Bagnulo}, {Sterzik}, \&
  {Fossati}}]{Bagnulo2011}
{Bagnulo}, S., {Sterzik}, M., \& {Fossati}, L. 2011, in Astronomical Society of
  the Pacific Conference Series, Vol. 449, Astronomical Polarimetry 2008:
  Science from Small to Large Telescopes, ed. P.~{Bastien}, N.~{Manset}, D.~P.
  {Clemens}, \& N.~{St-Louis}, 76

\bibitem[{{Blandford} {et~al.}(2019){Blandford}, {Meier}, \&
  {Readhead}}]{Blandford2019}
{Blandford}, R., {Meier}, D., \& {Readhead}, A. 2019, \araa, 57, 467

\bibitem[{{Blinov} {et~al.}(2021){Blinov}, {Kiehlmann}, {Pavlidou},
  {Panopoulou}, {Skalidis}, {Angelakis}, {Casadio}, {Einoder}, {Hovatta},
  {Kokolakis}, {Kougentakis}, {Kus}, {Kylafis}, {Kyritsis}, {Lalakos},
  {Liodakis}, {Maharana}, {Makrydopoulou}, {Mandarakas}, {Maragkakis},
  {Myserlis}, {Papadakis}, {Paterakis}, {Pearson}, {Ramaprakash}, {Readhead},
  {Reig}, {S{\l}owikowska}, {Tassis}, {Xexakis}, {{\.Z}ejmo}, \&
  {Zensus}}]{Blinov2021}
{Blinov}, D., {Kiehlmann}, S., {Pavlidou}, V., {et~al.} 2021, \mnras, 501, 3715

\bibitem[{{Cerruti}(2020)}]{Cerruti2020}
{Cerruti}, M. 2020, Galaxies, 8, 72

\bibitem[{{de Jaeger} {et~al.}(2023){de Jaeger}, {Shappee}, {Kochanek},
  {Hinkle}, {Garrappa}, {Liodakis}, {Franckowiak}, {Stanek}, {Beacom}, \&
  {Prieto}}]{deJaeger2023}
{de Jaeger}, T., {Shappee}, B.~J., {Kochanek}, C.~S., {et~al.} 2023, \mnras,
  519, 6349

\bibitem[{{Di Gesu} {et~al.}(2022){Di Gesu}, {Donnarumma}, {Tavecchio},
  {Agudo}, {Barnounin}, {Cibrario}, {Di Lalla}, {Di Marco}, {Escudero},
  {Errando}, {Jorstad}, {Kim}, {Kouch}, {Liodakis}, {Lindfors}, {Madejski},
  {Marshall}, {Marscher}, {Middei}, {Muleri}, {Myserlis}, {Negro}, {Omodei},
  {Pacciani}, {Paggi}, {Perri}, {Puccetti}, {Antonelli}, {Bachetti}, {Baldini},
  {Baumgartner}, {Bellazzini}, {Bianchi}, {Bongiorno}, {Bonino}, {Brez},
  {Bucciantini}, {Capitanio}, {Castellano}, {Cavazzuti}, {Ciprini}, {Costa},
  {De Rosa}, {Del Monte}, {Doroshenko}, {Dov{\v{c}}iak}, {Ehlert}, {Enoto},
  {Evangelista}, {Fabiani}, {Ferrazzoli}, {Garcia}, {Gunji}, {Hayashida},
  {Heyl}, {Iwakiri}, {Karas}, {Kitaguchi}, {Kolodziejczak}, {Krawczynski}, {La
  Monaca}, {Latronico}, {Maldera}, {Manfreda}, {Marin}, {Marinucci}, {Massaro},
  {Matt}, {Mitsuishi}, {Mizuno}, {Ng}, {O'Dell}, {Oppedisano}, {Papitto},
  {Pavlov}, {Peirson}, {Pesce-Rollins}, {Petrucci}, {Pilia}, {Possenti},
  {Poutanen}, {Ramsey}, {Rankin}, {Ratheesh}, {Romani}, {Sgr{\`o}}, {Slane},
  {Soffitta}, {Spandre}, {Tamagawa}, {Taverna}, {Tawara}, {Tennant}, {Thomas},
  {Tombesi}, {Trois}, {Tsygankov}, {Turolla}, {Vink}, {Weisskopf}, {Wu}, {Xie},
  \& {Zane}}]{DiGesu2022}
{Di Gesu}, L., {Donnarumma}, I., {Tavecchio}, F., {et~al.} 2022, \apjl, 938, L7

\bibitem[{{Di Gesu} {et~al.}(2023){Di Gesu}, {Marshall}, {Ehlert}, {Kim},
  {Donnarumma}, {Tavecchio}, {Liodakis}, {Kiehlmann}, {Agudo}, {Jorstad},
  {Muleri}, {Marscher}, {Puccetti}, {Middei}, {Perri}, {Pacciani}, {Negro},
  {Romani}, {Di Marco}, {Blinov}, {Bourbah}, {Kontopodis}, {Mandarakas},
  {Romanopoulos}, {Skalidis}, {Vervelaki}, {Casadio}, {Escudero}, {Myserlis},
  {Gurwell}, {Rao}, {Keating}, {Kouch}, {Lindfors}, {Aceituno}, {Bernardos},
  {Bonnoli}, {Casanova}, {Garc{\'\i}a-Comas}, {Ag{\'\i}s-Gonz{\'a}lez},
  {Husillos}, {Marchini}, {Sota}, {Imazawa}, {Sasada}, {Fukazawa}, {Kawabata},
  {Uemura}, {Mizuno}, {Nakaoka}, {Akitaya}, {Savchenko}, {Vasilyev},
  {G{\'o}mez}, {Antonelli}, {Barnouin}, {Bonino}, {Cavazzuti}, {Costamante},
  {Chen}, {Cibrario}, {De Rosa}, {Di Pierro}, {Errando}, {Kaaret}, {Karas},
  {Krawczynski}, {Lisalda}, {Madejski}, {Malacaria}, {Marin}, {Marinucci},
  {Massaro}, {Matt}, {Mitsuishi}, {O'Dell}, {Paggi}, {Peirson}, {Petrucci},
  {Ramsey}, {Tennant}, {Wu}, {Bachetti}, {Baldini}, {Baumgartner},
  {Bellazzini}, {Bianchi}, {Bongiorno}, {Brez}, {Bucciantini}, {Capitanio},
  {Castellano}, {Ciprini}, {Costa}, {Del Monte}, {Di Lalla}, {Doroshenko},
  {Dov{\v{c}}iak}, {Enoto}, {Evangelista}, {Fabiani}, {Ferrazzoli}, {Garcia},
  {Gunji}, {Hayashida}, {Heyl}, {Iwakiri}, {Kislat}, {Kitaguchi},
  {Kolodziejczak}, {La Monaca}, {Latronico}, {Maldera}, {Manfreda}, {Ng},
  {Omodei}, {Oppedisano}, {Papitto}, {Pavlov}, {Pesce-Rollins}, {Pilia},
  {Possenti}, {Poutanen}, {Rankin}, {Ratheesh}, {Roberts}, {Sgr{\`o}}, {Slane},
  {Soffitta}, {Spandre}, {Swartz}, {Tamagawa}, {Taverna}, {Tawara}, {Thomas},
  {Tombesi}, {Trois}, {Tsygankov}, {Turolla}, {Vink}, {Weisskopf}, {Xie}, \&
  {Zane}}]{DiGesu2023}
{Di Gesu}, L., {Marshall}, H.~L., {Ehlert}, S.~R., {et~al.} 2023, Nature
  Astronomy [\eprint[arXiv]{2305.13497}]

\bibitem[{{Georganopoulos} \& {Kazanas}(2003)}]{Georganopoulos2003}
{Georganopoulos}, M. \& {Kazanas}, D. 2003, \apjl, 594, L27

\bibitem[{{Hodge} {et~al.}(2018){Hodge}, {Lister}, {Aller}, {Aller}, {Kovalev},
  {Pushkarev}, \& {Savolainen}}]{Hodge2018}
{Hodge}, M.~A., {Lister}, M.~L., {Aller}, M.~F., {et~al.} 2018, \apj, 862, 151

\bibitem[{{Hovatta} \& {Lindfors}(2019)}]{Hovatta2019}
{Hovatta}, T. \& {Lindfors}, E. 2019, \nar, 87, 101541

\bibitem[{{Hovatta} {et~al.}(2016){Hovatta}, {Lindfors}, {Blinov}, {Pavlidou},
  {Nilsson}, {Kiehlmann}, {Angelakis}, {Fallah Ramazani}, {Liodakis},
  {Myserlis}, {Panopoulou}, \& {Pursimo}}]{Hovatta2016}
{Hovatta}, T., {Lindfors}, E., {Blinov}, D., {et~al.} 2016, \aap, 596, A78

\bibitem[{{Hovatta} {et~al.}(2021){Hovatta}, {Lindfors}, {Kiehlmann},
  {Max-Moerbeck}, {Hodges}, {Liodakis}, {L{\"a}hteem{\"a}ki}, {Pearson},
  {Readhead}, {Reeves}, {Suutarinen}, {Tammi}, \& {Tornikoski}}]{Hovatta2021}
{Hovatta}, T., {Lindfors}, E., {Kiehlmann}, S., {et~al.} 2021, \aap, 650, A83

\bibitem[{{Huppenkothen} {et~al.}(2015){Huppenkothen}, {Brewer}, {Hogg},
  {Murray}, {Frean}, {Elenbaas}, {Watts}, {Levin}, {van der Horst}, \&
  {Kouveliotou}}]{Huppenkothen2015}
{Huppenkothen}, D., {Brewer}, B.~J., {Hogg}, D.~W., {et~al.} 2015, \apj, 810,
  66

\bibitem[{{Hutsem{\'e}kers} {et~al.}(2010){Hutsem{\'e}kers}, {Borguet},
  {Sluse}, {Cabanac}, \& {Lamy}}]{Hutsemekers2010}
{Hutsem{\'e}kers}, D., {Borguet}, B., {Sluse}, D., {Cabanac}, R., \& {Lamy}, H.
  2010, \aap, 520, L7

\bibitem[{{IceCube Collaboration} {et~al.}(2018){IceCube Collaboration},
  {Aartsen}, {Ackermann}, {Adams}, {Aguilar}, {Ahlers}, {Ahrens}, {Samarai},
  {Altmann}, {Andeen}, {Anderson}, {Ansseau}, {Anton}, {Arg{\"u}elles},
  {Arsioli}, {Auffenberg}, {Axani}, {Bagherpour}, {Bai}, {Barron}, {Barwick},
  {Baum}, {Bay}, {Beatty}, {Becker Tjus}, {Becker}, {BenZvi}, {Berley},
  {Bernardini}, {Besson}, {Binder}, {Bindig}, {Blaufuss}, {Blot}, {Bohm},
  {B{\"o}rner}, {Bos}, {B{\"o}ser}, {Botner}, {Bourbeau}, {Bourbeau},
  {Bradascio}, {Braun}, {Brenzke}, {Bretz}, {Bron}, {Brostean-Kaiser},
  {Burgman}, {Busse}, {Carver}, {Cheung}, {Chirkin}, {Christov}, {Clark},
  {Classen}, {Coenders}, {Collin}, {Conrad}, {Coppin}, {Correa}, {Cowen},
  {Cross}, {Dave}, {Day}, {de Andr{\'e}}, {De Clercq}, {DeLaunay}, {Dembinski},
  {DeRidder}, {Desiati}, {de Vries}, {de Wasseige}, {de With}, {DeYoung},
  {D{\'\i}az-V{\'e}lez}, {di Lorenzo}, {Dujmovic}, {Dumm}, {Dunkman}, {Dvorak},
  {Eberhardt}, {Ehrhardt}, {Eichmann}, {Eller}, {Evenson}, {Fahey}, {Fazely},
  {Felde}, {Filimonov}, {Finley}, {Flis}, {Franckowiak}, {Friedman}, {Fritz},
  {Gaisser}, {Gallagher}, {Gerhardt}, {Ghorbani}, {Giommi}, {Glauch},
  {Gl{\"u}senkamp}, {Goldschmidt}, {Gonzalez}, {Grant}, {Griffith}, {Haack},
  {Hallgren}, {Halzen}, {Hanson}, {Hebecker}, {Heereman}, {Helbing},
  {Hellauer}, {Hickford}, {Hignight}, {Hill}, {Hoffman}, {Hoffmann}, {Hoinka},
  {Hokanson-Fasig}, {Hoshina}, {Huang}, {Huber}, {Hultqvist}, {H{\"u}nnefeld},
  {Hussain}, {In}, {Iovine}, {Ishihara}, {Jacobi}, {Japaridze}, {Jeong},
  {Jero}, {Jones}, {Kalaczynski}, {Kang}, {Kappes}, {Kappesser}, {Karg},
  {Karle}, {Katz}, {Kauer}, {Keivani}, {Kelley}, {Kheirandish}, {Kim}, {Kim},
  {Kintscher}, {Kiryluk}, {Kittler}, {Klein}, {Koirala}, {Kolanoski},
  {K{\"o}pke}, {Kopper}, {Kopper}, {Koschinsky}, {Koskinen}, {Kowalski},
  {Krammer}, {Krings}, {Kroll}, {Kr{\"u}ckl}, {Kunwar}, {Kurahashi},
  {Kuwabara}, {Kyriacou}, {Labare}, {Lanfranchi}, {Larson}, {Lauber},
  {Leonard}, {Lesiak-Bzdak}, {Leuermann}, {Liu}, {Lozano Mariscal}, {Lu},
  {L{\"u}nemann}, {Luszczak}, {Madsen}, {Maggi}, {Mahn}, {Mancina}, {Maruyama},
  {Mase}, {Maunu}, {Meagher}, {Medici}, {Meier}, {Menne}, {Merino}, {Meures},
  {Miarecki}, {Micallef}, {Moment{\'e}}, {Montaruli}, {Moore}, {Morse},
  {Moulai}, {Nahnhauer}, {Nakarmi}, {Naumann}, {Neer}, {Niederhausen},
  {Nowicki}, {Nygren}, {Obertacke Pollmann}, {Olivas}, {O'Murchadha},
  {O'Sullivan}, {Padovani}, {Palczewski}, {Pandya}, {Pankova}, {Peiffer},
  {Pepper}, {P{\'e}rez de los Heros}, {Pieloth}, {Pinat}, {Plum}, {Price},
  {Przybylski}, {Raab}, {R{\"a}del}, {Rameez}, {Rawlins}, {Rea}, {Reimann},
  {Relethford}, {Relich}, {Resconi}, {Rhode}, {Richman}, {Robertson}, {Rongen},
  {Rott}, {Ruhe}, {Ryckbosch}, {Rysewyk}, {Safa}, {Sahakyan}, {S{\"a}lzer},
  {Sanchez Herrera}, {Sandrock}, {Sandroos}, {Santander}, {Sarkar}, {Sarkar},
  {Satalecka}, {Schlunder}, {Schmidt}, {Schneider}, {Schoenen},
  {Sch{\"o}neberg}, {Schumacher}, {Sclafani}, {Seckel}, {Seunarine},
  {Soedingrekso}, {Soldin}, {Song}, {Spiczak}, {Spiering}, {Stachurska},
  {Stamatikos}, {Stanev}, {Stasik}, {Stettner}, {Steuer}, {Stezelberger},
  {Stokstad}, {St{\"o}{\ss}l}, {Strotjohann}, {Stuttard}, {Sullivan},
  {Sutherland}, {Taboada}, {Tatar}, {Tenholt}, {Ter-Antonyan}, {Terliuk},
  {Tilav}, {Toale}, {Tobin}, {Toennis}, {Toscano}, {Tosi}, {Tselengidou},
  {Tung}, {Turcati}, {Turley}, {Ty}, {Unger}, {Usner}, {Vandenbroucke}, {Van
  Driessche}, {van Eijk}, {van Eijndhoven}, {Vanheule}, {van Santen}, {Vogel},
  {Vraeghe}, {Walck}, {Wallace}, {Wallraff}, {Wandler}, {Wandkowsky}, {Waza},
  {Weaver}, {Weiss}, {Wendt}, {Werthebach}, {Westerhoff}, {Whelan},
  {Whitehorn}, {Wiebe}, {Wiebusch}, {Wille}, {Williams}, {Wills}, {Wolf},
  {Wood}, {Wood}, {Woschnagg}, {Xu}, {Xu}, {Xu}, {Yanez}, {Yodh}, {Yoshida}, \&
  {Yuan}}]{icecube2018}
{IceCube Collaboration}, {Aartsen}, M.~G., {Ackermann}, M., {et~al.} 2018,
  Science, 361, 147

\bibitem[{{Jones} \& {O'Dell}(1977)}]{Jones1977}
{Jones}, T.~W. \& {O'Dell}, S.~L. 1977, \apj, 214, 522

\bibitem[{{Kiehlmann} {et~al.}(2021){Kiehlmann}, {Blinov}, {Liodakis},
  {Pavlidou}, {Readhead}, {Angelakis}, {Casadio}, {Hovatta}, {Kylafis},
  {Mahabal}, {Mandarakas}, {Myserlis}, {Panopoulou}, {Pearson}, {Ramaprakash},
  {Reig}, {Skalidis}, {S{\l}owikowska}, {Tassis}, \& {Zensus}}]{Kiehlmann2021}
{Kiehlmann}, S., {Blinov}, D., {Liodakis}, I., {et~al.} 2021, \mnras, 507, 225

\bibitem[{{Kovalev} {et~al.}(2023){Kovalev}, {Plavin}, {Pushkarev}, \&
  {Troitsky}}]{Kovalev2023}
{Kovalev}, Y.~Y., {Plavin}, A.~V., {Pushkarev}, A.~B., \& {Troitsky}, S.~V.
  2023, Galaxies, 11, 84

\bibitem[{{Lico} {et~al.}(2020){Lico}, {Liu}, {Giroletti}, {Orienti},
  {G{\'o}mez}, {Piner}, {MacDonald}, {D'Ammando}, \& {Fuentes}}]{Lico2020}
{Lico}, R., {Liu}, J., {Giroletti}, M., {et~al.} 2020, \aap, 634, A87

\bibitem[{{Liodakis} {et~al.}(2020){Liodakis}, {Blinov}, {Jorstad}, {Arkharov},
  {Di Paola}, {Efimova}, {Grishina}, {Kiehlmann}, {Kopatskaya}, {Larionov},
  {Larionova}, {Larionova}, {Marscher}, {Morozova}, {Nikiforova}, {Pavlidou},
  {Traianou}, {Troitskaya}, {Troitsky}, {Uemura}, \& {Weaver}}]{Liodakis2020}
{Liodakis}, I., {Blinov}, D., {Jorstad}, S.~G., {et~al.} 2020, \apj, 902, 61

\bibitem[{{Liodakis} {et~al.}(2022{\natexlab{a}}){Liodakis}, {Blinov},
  {Potter}, \& {Rieger}}]{Liodakis2022-II}
{Liodakis}, I., {Blinov}, D., {Potter}, S.~B., \& {Rieger}, F.~M.
  2022{\natexlab{a}}, \mnras, 509, L21

\bibitem[{{Liodakis} {et~al.}(2021){Liodakis}, {Hovatta}, {Aller}, {Aller},
  {Gurwell}, {L{\"a}hteenm{\"a}ki}, \& {Tornikoski}}]{Liodakis2021}
{Liodakis}, I., {Hovatta}, T., {Aller}, M.~F., {et~al.} 2021, \aap, 654, A169

\bibitem[{{Liodakis} {et~al.}(2018){Liodakis}, {Hovatta}, {Huppenkothen},
  {Kiehlmann}, {Max-Moerbeck}, \& {Readhead}}]{Liodakis2018}
{Liodakis}, I., {Hovatta}, T., {Huppenkothen}, D., {et~al.} 2018, \apj, 866,
  137

\bibitem[{{Liodakis} {et~al.}(2017){Liodakis}, {Marchili}, {Angelakis},
  {Fuhrmann}, {Nestoras}, {Myserlis}, {Karamanavis}, {Krichbaum}, {Sievers},
  {Ungerechts}, \& {Zensus}}]{Liodakis2017}
{Liodakis}, I., {Marchili}, N., {Angelakis}, E., {et~al.} 2017, \mnras, 466,
  4625

\bibitem[{{Liodakis} {et~al.}(2022{\natexlab{b}}){Liodakis}, {Marscher},
  {Agudo}, {Berdyugin}, {Bernardos}, {Bonnoli}, {Borman}, {Casadio},
  {Casanova}, {Cavazzuti}, {Rodriguez Cavero}, {Di Gesu}, {Di Lalla},
  {Donnarumma}, {Ehlert}, {Errando}, {Escudero}, {Garc{\'\i}a-Comas},
  {Ag{\'\i}s-Gonz{\'a}lez}, {Husillos}, {Jormanainen}, {Jorstad}, {Kagitani},
  {Kopatskaya}, {Kravtsov}, {Krawczynski}, {Lindfors}, {Larionova}, {Madejski},
  {Marin}, {Marchini}, {Marshall}, {Morozova}, {Massaro}, {Masiero}, {Mawet},
  {Middei}, {Millar-Blanchaer}, {Myserlis}, {Negro}, {Nilsson}, {O'Dell},
  {Omodei}, {Pacciani}, {Paggi}, {Panopoulou}, {Peirson}, {Perri}, {Petrucci},
  {Poutanen}, {Puccetti}, {Romani}, {Sakanoi}, {Savchenko}, {Sota},
  {Tavecchio}, {Tinyanont}, {Vasilyev}, {Weaver}, {Zhovtan}, {Antonelli},
  {Bachetti}, {Baldini}, {Baumgartner}, {Bellazzini}, {Bianchi}, {Bongiorno},
  {Bonino}, {Brez}, {Bucciantini}, {Capitanio}, {Castellano}, {Ciprini},
  {Costa}, {De Rosa}, {Del Monte}, {Di Marco}, {Doroshenko}, {Dov{\v{c}}iak},
  {Enoto}, {Evangelista}, {Fabiani}, {Ferrazzoli}, {Garcia}, {Gunji},
  {Hayashida}, {Heyl}, {Iwakiri}, {Karas}, {Kitaguchi}, {Kolodziejczak}, {La
  Monaca}, {Latronico}, {Maldera}, {Manfreda}, {Marinucci}, {Matt},
  {Mitsuishi}, {Mizuno}, {Muleri}, {Ng}, {Oppedisano}, {Papitto}, {Pavlov},
  {Pesce-Rollins}, {Pilia}, {Possenti}, {Ramsey}, {Rankin}, {Ratheesh},
  {Sgr{\'o}}, {Slane}, {Soffitta}, {Spandre}, {Tamagawa}, {Taverna}, {Tawara},
  {Tennant}, {Thomas}, {Tombesi}, {Trois}, {Tsygankov}, {Turolla}, {Vink},
  {Weisskopf}, {Wu}, {Xie}, \& {Zane}}]{Liodakis2022}
{Liodakis}, I., {Marscher}, A.~P., {Agudo}, I., {et~al.} 2022{\natexlab{b}},
  \nat, 611, 677

\bibitem[{{Liodakis} \& {Petropoulou}(2020)}]{Liodakis2020-II}
{Liodakis}, I. \& {Petropoulou}, M. 2020, \apjl, 893, L20

\bibitem[{{Liodakis} {et~al.}(2019){Liodakis}, {Romani}, {Filippenko},
  {Kocevski}, \& {Zheng}}]{Liodakis2019}
{Liodakis}, I., {Romani}, R.~W., {Filippenko}, A.~V., {Kocevski}, D., \&
  {Zheng}, W. 2019, \apj, 880, 32

\bibitem[{{MAGIC Collaboration} {et~al.}(2023){MAGIC Collaboration}, {Acciari},
  {Aniello}, {Ansoldi}, {Antonelli}, {Arbet Engels}, {Arcaro}, {Artero},
  {Asano}, {Baack}, {Babi{\'c}}, {Baquero}, {Barres de Almeida}, {Barrio},
  {Batkovi{\'c}}, {Becerra Gonz{\'a}lez}, {Bednarek}, {Bernardini},
  {Bernardos}, {Berti}, {Besenrieder}, {Bhattacharyya}, {Bigongiari}, {Biland},
  {Blanch}, {B{\"o}kenkamp}, {Bonnoli}, {Bo{\v{s}}njak}, {Burelli}, {Busetto},
  {Carosi}, {Carretero-Castrillo}, {Ceribella}, {Chai}, {Chilingarian},
  {Cikota}, {Colombo}, {Contreras}, {Cortina}, {Covino}, {D'Amico}, {D'Elia},
  {da Vela}, {Dazzi}, {de Angelis}, {de Lotto}, {Del Popolo}, {Delfino},
  {Delgado}, {Delgado Mendez}, {Depaoli}, {di Pierro}, {di Venere}, {Do Souto
  Espi{\~n}eira}, {Dominis Prester}, {Donini}, {Dorner}, {Doro}, {Elsaesser},
  {Emery}, {Fallah Ramazani}, {Fari{\~n}a}, {Fattorini}, {Font}, {Fruck},
  {Fukami}, {Fukazawa}, {Garc{\'\i}a L{\'o}pez}, {Garczarczyk}, {Gasparyan},
  {Gaug}, {Giesbrecht Paiva}, {Giglietto}, {Giordano}, {Gliwny},
  {Godinovi{\'c}}, {Green}, {Green}, {Hadasch}, {Hahn}, {Hassan}, {Heckmann},
  {Herrera}, {Hrupec}, {H{\"u}tten}, {Inada}, {Iotov}, {Ishio}, {Iwamura},
  {Jim{\'e}nez Mart{\'\i}nez}, {Jormanainen}, {Kerszberg}, {Kobayashi}, {Kubo},
  {Kushida}, {Lamastra}, {Lelas}, {Leone}, {Lindfors}, {Linhoff}, {Lombardi},
  {Longo}, {L{\'o}pez-Coto}, {L{\'o}pez-Moya}, {L{\'o}pez-Oramas}, {Loporchio},
  {Lorini}, {Lyard}, {Machado de Oliveira Fraga}, {Majumdar}, {Makariev},
  {Maneva}, {Manganaro}, {Mangano}, {Mannheim}, {Mariotti}, {Mart{\'\i}nez},
  {Mas Aguilar}, {Mazin}, {Menchiari}, {Mender}, {Mi{\'c}anovi{\'c}}, {Miceli},
  {Miener}, {Miranda}, {Mirzoyan}, {Molina}, {Mondal}, {Moralejo}, {Morcuende},
  {Moreno}, {Nakamori}, {Nanci}, {Nava}, {Neustroev}, {Nievas Rosillo},
  {Nigro}, {Nilsson}, {Nishijima}, {Njoh Ekoume}, {Noda}, {Nozaki}, {Ohtani},
  {Oka}, {Otero-Santos}, {Paiano}, {Palatiello}, {Paneque}, {Paoletti},
  {Paredes}, {Pavleti{\'c}}, {Persic}, {Pihet}, {Podobnik}, {Prada Moroni},
  {Prandini}, {Principe}, {Priyadarshi}, {Puljak}, {Rhode}, {Rib{\'o}}, {Rico},
  {Righi}, {Rugliancich}, {Sahakyan}, {Saito}, {Sakurai}, {Satalecka},
  {Saturni}, {Schleicher}, {Schmidt}, {Schmuckermaier}, {Schubert},
  {Schweizer}, {Sitarek}, {Sliusar}, {Sobczynska}, {Spolon}, {Stamerra},
  {Stri{\v{s}}kovi{\'c}}, {Strom}, {Strzys}, {Suda}, {Suri{\'c}}, {Takahashi},
  {Takeishi}, {Tavecchio}, {Temnikov}, {Terzi{\'c}}, {Teshima}, {Tosti},
  {Truzzi}, {Tutone}, {Ubach}, {van Scherpenberg}, {Vanzo}, {Vazquez Acosta},
  {Ventura}, {Verguilov}, {Viale}, {Vigorito}, {Vitale}, {Vovk}, {Walter},
  {Will}, {Wunderlich}, {Yamamoto}, {Zari{\'c}}, {Acosta-Pulido}, {D'Ammando},
  {Hovatta}, {Kiehlmann}, {Liodakis}, {Leto}, {Max-Moerbeck}, {Pacciani},
  {Perri}, {Readhead}, {Reeves}, \& {Verrecchia}}]{magic2023}
{MAGIC Collaboration}, {Acciari}, V.~A., {Aniello}, T., {et~al.} 2023, \aap,
  670, A49

\bibitem[{{MAGIC Collaboration} {et~al.}(2018){MAGIC Collaboration}, {Ahnen},
  {Ansoldi}, {Antonelli}, {Arcaro}, {Baack}, {Babi{\'c}}, {Banerjee},
  {Bangale}, {Barres de Almeida}, {Barrio}, {Bednarek}, {Bernardini}, {Berse},
  {Berti}, {Bhattacharyya}, {Biland}, {Blanch}, {Bonnoli}, {Carosi}, {Carosi},
  {Ceribella}, {Chatterjee}, {Colak}, {Colin}, {Colombo}, {Contreras},
  {Cortina}, {Covino}, {Cumani}, {da Vela}, {Dazzi}, {de Angelis}, {de Lotto},
  {Delfino}, {Delgado}, {di Pierro}, {Dom{\'\i}nguez}, {Dominis Prester},
  {Dorner}, {Doro}, {Einecke}, {Elsaesser}, {Fallah Ramazani},
  {Fern{\'a}ndez-Barral}, {Fidalgo}, {Fonseca}, {Font}, {Fruck}, {Galindo},
  {Garc{\'\i}a L{\'o}pez}, {Garczarczyk}, {Gaug}, {Giammaria}, {Godinovi{\'c}},
  {Gora}, {Guberman}, {Hadasch}, {Hahn}, {Hassan}, {Hayashida}, {Herrera},
  {Hose}, {Hrupec}, {Ishio}, {Konno}, {Kubo}, {Kushida}, {Kuve{\v{z}}di{\'c}},
  {Lelas}, {Lindfors}, {Lombardi}, {Longo}, {L{\'o}pez}, {Maggio}, {Majumdar},
  {Makariev}, {Maneva}, {Manganaro}, {Mannheim}, {Maraschi}, {Mariotti},
  {Mart{\'\i}nez}, {Masuda}, {Mazin}, {Mielke}, {Minev}, {Miranda}, {Mirzoyan},
  {Moralejo}, {Moreno}, {Moretti}, {Nagayoshi}, {Neustroev}, {Niedzwiecki},
  {Nievas Rosillo}, {Nigro}, {Nilsson}, {Ninci}, {Nishijima}, {Noda},
  {Nogu{\'e}s}, {Paiano}, {Palacio}, {Paneque}, {Paoletti}, {Paredes},
  {Pedaletti}, {Peresano}, {Persic}, {Prada Moroni}, {Prandini}, {Puljak},
  {Garcia}, {Reichardt}, {Rhode}, {Rib{\'o}}, {Rico}, {Righi}, {Rugliancich},
  {Saito}, {Satalecka}, {Schweizer}, {Sitarek}, {{\v{S}}nidari{\'c}},
  {Sobczynska}, {Stamerra}, {Strzys}, {Suri{\'c}}, {Takahashi}, {Takalo},
  {Tavecchio}, {Temnikov}, {Terzi{\'c}}, {Teshima}, {Torres-Alb{\`a}},
  {Treves}, {Tsujimoto}, {Vanzo}, {Vazquez Acosta}, {Vovk}, {Ward}, {Will},
  {Zari{\'c}}, {Becerra Gonz{\'a}lez}, {Tanaka}, {Ojha}, {Finke},
  {L{\"a}hteenm{\"a}ki}, {J{\"a}rvel{\"a}}, {Tornikoski}, {Ramakrishnan},
  {Hovatta}, {Jorstad}, {Marscher}, {Larionov}, {Borman}, {Grishina},
  {Kopatskaya}, {Larionova}, {Morozova}, {Savchenko}, {Troitskaya}, {Troitsky},
  {Vasilyev}, {Agudo}, {Molina}, {Casadio}, {Gurwell}, {Carnerero}, {Protasio},
  \& {Acosta Pulido}}]{MAGIC2018}
{MAGIC Collaboration}, {Ahnen}, M.~L., {Ansoldi}, S., {et~al.} 2018, \aap, 617,
  A30

\bibitem[{{Marscher} \& {Gear}(1985)}]{Marscher1985}
{Marscher}, A.~P. \& {Gear}, W.~K. 1985, \apj, 298, 114

\bibitem[{{Middei} {et~al.}(2023{\natexlab{a}}){Middei}, {Liodakis}, {Perri},
  {Puccetti}, {Cavazzuti}, {Di Gesu}, {Ehlert}, {Madejski}, {Marscher},
  {Marshall}, {Muleri}, {Negro}, {Jorstad}, {Ag{\'\i}s-Gonz{\'a}lez}, {Agudo},
  {Bonnoli}, {Bernardos}, {Casanova}, {Garc{\'\i}a-Comas}, {Husillos},
  {Marchini}, {Sota}, {Kouch}, {Lindfors}, {Borman}, {Kopatskaya}, {Larionova},
  {Morozova}, {Savchenko}, {Vasilyev}, {Zhovtan}, {Casadio}, {Escudero},
  {Myserlis}, {Hales}, {Kameno}, {Kneissl}, {Messias}, {Nagai}, {Blinov},
  {Bourbah}, {Kiehlmann}, {Kontopodis}, {Mandarakas}, {Romanopoulos},
  {Skalidis}, {Vervelaki}, {Masiero}, {Mawet}, {Millar-Blanchaer},
  {Panopoulou}, {Tinyanont}, {Berdyugin}, {Kagitani}, {Kravtsov}, {Sakanoi},
  {Imazawa}, {Sasada}, {Fukazawa}, {Kawabata}, {Uemura}, {Mizuno}, {Nakaoka},
  {Akitaya}, {Gurwell}, {Rao}, {Di Lalla}, {Cibrario}, {Donnarumma}, {Kim},
  {Omodei}, {Pacciani}, {Poutanen}, {Tavecchio}, {Antonelli}, {Bachetti},
  {Baldini}, {Baumgartner}, {Bellazzini}, {Bianchi}, {Bongiorno}, {Bonino},
  {Brez}, {Bucciantini}, {Capitanio}, {Castellano}, {Ciprini}, {Costa}, {De
  Rosa}, {Del Monte}, {Di Marco}, {Doroshenko}, {Dov{\v{c}}iak}, {Enoto},
  {Evangelista}, {Fabiani}, {Ferrazzoli}, {Garcia}, {Gunji}, {Hayashida},
  {Heyl}, {Iwakiri}, {Karas}, {Kitaguchi}, {Kolodziejczak}, {Krawczynski}, {La
  Monaca}, {Latronico}, {Maldera}, {Manfreda}, {Marin}, {Marinucci}, {Massaro},
  {Matt}, {Mitsuishi}, {Ng}, {O'Dell}, {Oppedisano}, {Papitto}, {Pavlov},
  {Peirson}, {Pesce-Rollins}, {Petrucci}, {Pilia}, {Possenti}, {Ramsey},
  {Rankin}, {Ratheesh}, {Romani}, {Sgr{\'o}}, {Slane}, {Soffitta}, {Spandre},
  {Tamagawa}, {Taverna}, {Tawara}, {Tennant}, {Thomas}, {Tombesi}, {Trois},
  {Tsygankov}, {Turolla}, {Vink}, {Weisskopf}, {Wu}, {Xie}, \&
  {Zane}}]{Middei2023}
{Middei}, R., {Liodakis}, I., {Perri}, M., {et~al.} 2023{\natexlab{a}}, \apjl,
  942, L10

\bibitem[{{Middei} {et~al.}(2023{\natexlab{b}}){Middei}, {Perri}, {Puccetti},
  {Liodakis}, {Di Gesu}, {Marscher}, {Rodriguez Cavero}, {Tavecchio},
  {Donnarumma}, {Laurenti}, {Jorstad}, {Agudo}, {Marshall}, {Pacciani}, {Kim},
  {Aceituno}, {Bonnoli}, {Casanova}, {Ag{\'\i}s-Gonz{\'a}lez}, {Sota},
  {Casadio}, {Escudero}, {Myserlis}, {Sievers}, {Kouch}, {Lindfors}, {Gurwell},
  {Keating}, {Rao}, {Kang}, {Lee}, {Kim}, {Cheong}, {Jeong}, {Angelakis},
  {Kraus}, {Antonelli}, {Bachetti}, {Baldini}, {Baumgartner}, {Bellazzini},
  {Bianchi}, {Bongiorno}, {Bonino}, {Brez}, {Bucciantini}, {Capitanio},
  {Castellano}, {Cavazzuti}, {Chen}, {Ciprini}, {Costa}, {De Rosa}, {Del
  Monte}, {Di Lalla}, {Di Marco}, {Doroshenko}, {Dov{\v{c}}iak}, {Ehlert},
  {Enoto}, {Evangelista}, {Fabiani}, {Ferrazzoli}, {Garc{\'\i}a}, {Gunji},
  {Hayashida}, {Heyl}, {Iwakiri}, {Kaaret}, {Karas}, {Kislat}, {Kitaguchi},
  {Kolodziejczak}, {Krawczynski}, {La Monaca}, {Latronico}, {Maldera},
  {Manfreda}, {Marin}, {Marinucci}, {Massaro}, {Matt}, {Mitsuishi}, {Mizuno},
  {Muleri}, {Negro}, {Ng}, {O'Dell}, {Omodei}, {Oppedisano}, {Papitto},
  {Pavlov}, {Peirson}, {Pesce-Rollins}, {Petrucci}, {Pilia}, {Possenti},
  {Poutanen}, {Ramsey}, {Rankin}, {Ratheesh}, {Roberts}, {Romani}, {Sgr{\`o}},
  {Slane}, {Soffitta}, {Spandre}, {Swartz}, {Tamagawa}, {Taverna}, {Tawara},
  {Tennant}, {Thomas}, {Tombesi}, {Trois}, {Tsygankov}, {Turolla}, {Vink},
  {Weisskopf}, {Wu}, {Xie}, \& {Zane}}]{Middei2023-II}
{Middei}, R., {Perri}, M., {Puccetti}, S., {et~al.} 2023{\natexlab{b}}, \apjl,
  953, L28

\bibitem[{{Novikova} {et~al.}(2023){Novikova}, {Shishkina}, \&
  {Blinov}}]{Novikova2023}
{Novikova}, P., {Shishkina}, E., \& {Blinov}, D. 2023, \mnras
  [\eprint[arXiv]{2304.13044}]

\bibitem[{{Oikonomou} {et~al.}(2019){Oikonomou}, {Murase}, {Padovani},
  {Resconi}, \& {M{\'e}sz{\'a}ros}}]{Oikonomou2019}
{Oikonomou}, F., {Murase}, K., {Padovani}, P., {Resconi}, E., \&
  {M{\'e}sz{\'a}ros}, P. 2019, \mnras, 489, 4347

\bibitem[{{O'Sullivan} \& {Gabuzda}(2008)}]{osullivan2008}
{O'Sullivan}, S.~P. \& {Gabuzda}, D.~C. 2008, in Astronomical Society of the
  Pacific Conference Series, Vol. 386, Extragalactic Jets: Theory and
  Observation from Radio to Gamma Ray, ed. T.~A. {Rector} \& D.~S. {De Young},
  284

\bibitem[{{Peirson} {et~al.}(2022){Peirson}, {Liodakis}, \&
  {Romani}}]{Peirson2022}
{Peirson}, A.~L., {Liodakis}, I., \& {Romani}, R.~W. 2022, \apj, 931, 59

\bibitem[{{Peirson} {et~al.}(2023){Peirson}, {Negro}, {Liodakis}, {Middei},
  {Kim}, {Marscher}, {Marshall}, {Pacciani}, {Romani}, {Wu}, {Di Marco}, {Di
  Lalla}, {Omodei}, {Jorstad}, {Agudo}, {Kouch}, {Lindfors}, {Aceituno},
  {Bernardos}, {Bonnoli}, {Casanova}, {Garc{\'\i}a-Comas},
  {Ag{\'\i}s-Gonz{\'a}lez}, {Husillos}, {Marchini}, {Sota}, {Casadio},
  {Escudero}, {Myserlis}, {Sievers}, {Gurwell}, {Rao}, {Imazawa}, {Sasada},
  {Fukazawa}, {Kawabata}, {Uemura}, {Mizuno}, {Nakaoka}, {Akitaya}, {Cheong},
  {Jeong}, {Kang}, {Kim}, {Lee}, {Angelakis}, {Kraus}, {Cibrario},
  {Donnarumma}, {Poutanen}, {Tavecchio}, {Antonelli}, {Bachetti}, {Baldini},
  {Baumgartner}, {Bellazzini}, {Bianchi}, {Bongiorno}, {Bonino}, {Brez},
  {Bucciantini}, {Capitanio}, {Castellano}, {Cavazzuti}, {Chen}, {Ciprini},
  {Costa}, {De Rosa}, {Del Monte}, {Di Gesu}, {Doroshenko}, {Dov{\v{c}}iak},
  {Ehlert}, {Enoto}, {Evangelista}, {Fabiani}, {Ferrazzoli}, {Garcia}, {Gunji},
  {Hayashida}, {Heyl}, {Iwakiri}, {Kaaret}, {Karas}, {Kitaguchi},
  {Kolodziejczak}, {Krawczynski}, {La Monaca}, {Latronico}, {Madejski},
  {Maldera}, {Manfreda}, {Marin}, {Marinucci}, {Massaro}, {Matt}, {Mitsuishi},
  {Muleri}, {Ng}, {O'Dell}, {Oppedisano}, {Papitto}, {Pavlov}, {Perri},
  {Pesce-Rollins}, {Petrucci}, {Pilia}, {Possenti}, {Puccetti}, {Ramsey},
  {Rankin}, {Ratheesh}, {Roberts}, {Sgr{\'o}}, {Slane}, {Soffitta}, {Spandre},
  {Swartz}, {Tamagawa}, {Taverna}, {Tawara}, {Tennant}, {Thomas}, {Tombesi},
  {Trois}, {Tsygankov}, {Turolla}, {Vink}, {Weisskopf}, {Xie}, \&
  {Zane}}]{Peirson2023}
{Peirson}, A.~L., {Negro}, M., {Liodakis}, I., {et~al.} 2023, \apjl, 948, L25

\bibitem[{{Plavin} {et~al.}(2020){Plavin}, {Kovalev}, {Kovalev}, \&
  {Troitsky}}]{Plavin2020}
{Plavin}, A., {Kovalev}, Y.~Y., {Kovalev}, Y.~A., \& {Troitsky}, S. 2020, \apj,
  894, 101

\bibitem[{{Pushkarev} {et~al.}(2012){Pushkarev}, {Hovatta}, {Kovalev},
  {Lister}, {Lobanov}, {Savolainen}, \& {Zensus}}]{Pushkarev2012}
{Pushkarev}, A.~B., {Hovatta}, T., {Kovalev}, Y.~Y., {et~al.} 2012, \aap, 545,
  A113

\bibitem[{{Raiteri} {et~al.}(2017){Raiteri}, {Nicastro}, {Stamerra}, {Villata},
  {Larionov}, {Blinov}, {Acosta-Pulido}, {Ar{\'e}valo}, {Arkharov}, {Bachev},
  {Borman}, {Carnerero}, {Carosati}, {Cecconi}, {Chen}, {Damljanovic}, {Di
  Paola}, {Ehgamberdiev}, {Frasca}, {Giroletti}, {Gonz{\'a}lez-Morales},
  {Gri{\~n}on-Mar{\'\i}n}, {Grishina}, {Huang}, {Ibryamov}, {Klimanov},
  {Kopatskaya}, {Kurtanidze}, {Kurtanidze}, {L{\"a}hteenm{\"a}ki}, {Larionova},
  {Larionova}, {L{\'a}zaro}, {Leto}, {Liodakis}, {Mart{\'\i}nez-Lombilla},
  {Mihov}, {Mirzaqulov}, {Mokrushina}, {Moody}, {Morozova}, {Nazarov},
  {Nikolashvili}, {Ohlert}, {Panopoulou}, {Pastor Yabar}, {Pinna}, {Protasio},
  {Rizzi}, {Sadun}, {Savchenko}, {Semkov}, {Sigua}, {Slavcheva-Mihova},
  {Strigachev}, {Tornikoski}, {Troitskaya}, {Troitsky}, {Vasilyev}, {Vera},
  {Vince}, \& {Zanmar Sanchez}}]{Raiteri2017-II}
{Raiteri}, C.~M., {Nicastro}, F., {Stamerra}, A., {et~al.} 2017, \mnras, 466,
  3762

\bibitem[{{Raiteri} {et~al.}(2021){Raiteri}, {Villata}, {Larionov}, {Jorstad},
  {Marscher}, {Weaver}, {Acosta-Pulido}, {Agudo}, {Andreeva}, {Arkharov},
  {Bachev}, {Ben{\'\i}tez}, {Berton}, {Bj{\"o}rklund}, {Borman}, {Bozhilov},
  {Carnerero}, {Carosati}, {Casadio}, {Chen}, {Damljanovic}, {D'Ammando},
  {Escudero}, {Fuentes}, {Giroletti}, {Grishina}, {Gupta}, {Hagen-Thorn},
  {Hart}, {Hiriart}, {Hou}, {Ivanov}, {Kim}, {Kimeridze}, {Konstantopoulou},
  {Kopatskaya}, {Kurtanidze}, {Kurtanidze}, {L{\"a}hteenm{\"a}ki}, {Larionova},
  {Larionova}, {Marchili}, {Markovic}, {Minev}, {Morozova}, {Myserlis},
  {Nakamura}, {Nikiforova}, {Nikolashvili}, {Otero-Santos}, {Ovcharov},
  {Pursimo}, {Rahimov}, {Righini}, {Sakamoto}, {Savchenko}, {Semkov},
  {Shakhovskoy}, {Sigua}, {Stojanovic}, {Strigachev}, {Thum}, {Tornikoski},
  {Traianou}, {Troitskaya}, {Troitskiy}, {Tsai}, {Valcheva}, {Vasilyev},
  {Vince}, \& {Zaharieva}}]{Raiteri2021}
{Raiteri}, C.~M., {Villata}, M., {Larionov}, V.~M., {et~al.} 2021, \mnras, 504,
  5629

\bibitem[{{Richards} {et~al.}(2011){Richards}, {Max-Moerbeck}, {Pavlidou},
  {King}, {Pearson}, {Readhead}, {Reeves}, {Shepherd}, {Stevenson},
  {Weintraub}, {Fuhrmann}, {Angelakis}, {Zensus}, {Healey}, {Romani}, {Shaw},
  {Grainge}, {Birkinshaw}, {Lancaster}, {Worrall}, {Taylor}, {Cotter}, \&
  {Bustos}}]{Richards2011}
{Richards}, J.~L., {Max-Moerbeck}, W., {Pavlidou}, V., {et~al.} 2011, \apjs,
  194, 29

\bibitem[{{Rieger} \& {Mannheim}(2005)}]{Rieger2005}
{Rieger}, F.~M. \& {Mannheim}, K. 2005, Chinese Journal of Astronomy and
  Astrophysics Supplement, 5, 311

\bibitem[{{Sikora} {et~al.}(2009){Sikora}, {Stawarz}, {Moderski}, {Nalewajko},
  \& {Madejski}}]{Sikora2009}
{Sikora}, M., {Stawarz}, {\L}., {Moderski}, R., {Nalewajko}, K., \& {Madejski},
  G.~M. 2009, \apj, 704, 38

\bibitem[{{Stathopoulos} {et~al.}(2022){Stathopoulos}, {Petropoulou}, {Giommi},
  {Vasilopoulos}, {Padovani}, \& {Mastichiadis}}]{Stathopoulos2022}
{Stathopoulos}, S.~I., {Petropoulou}, M., {Giommi}, P., {et~al.} 2022, \mnras,
  510, 4063

\bibitem[{{Tanaka} {et~al.}(2016){Tanaka}, {Becerra Gonzalez}, {Itoh}, {Finke},
  {Inoue}, {Ojha}, {Carpenter}, {Lindfors}, {Krau{\ss}}, {Desiante}, {Shiki},
  {Fukazawa}, {Longo}, {McEnery}, {Buson}, {Nilsson}, {Fallah Ramazani},
  {Reinthal}, {Takalo}, {Pursimo}, \& {Boschin}}]{Tanaka2016}
{Tanaka}, Y.~T., {Becerra Gonzalez}, J., {Itoh}, R., {et~al.} 2016, \pasj, 68,
  51

\bibitem[{{Valtaoja} {et~al.}(1993){Valtaoja}, {Karttunen}, {Valtaoja},
  {Shakhovskoy}, \& {Efimov}}]{Valtaoja1993}
{Valtaoja}, L., {Karttunen}, H., {Valtaoja}, E., {Shakhovskoy}, N.~M., \&
  {Efimov}, Y.~S. 1993, \aap, 273, 393

\bibitem[{{Wagner} \& {Mannheim}(2001)}]{Wagner2001}
{Wagner}, S.~J. \& {Mannheim}, K. 2001, in Astronomical Society of the Pacific
  Conference Series, Vol. 250, Particles and Fields in Radio Galaxies
  Conference, ed. R.~A. {Laing} \& K.~M. {Blundell}, 142

\bibitem[{{Weaver} {et~al.}(2022){Weaver}, {Jorstad}, {Marscher}, {Morozova},
  {Troitsky}, {Agudo}, {G{\'o}mez}, {L{\"a}hteenm{\"a}ki}, {Tammi}, \&
  {Tornikoski}}]{Weaver2022}
{Weaver}, Z.~R., {Jorstad}, S.~G., {Marscher}, A.~P., {et~al.} 2022, \apjs,
  260, 12

\bibitem[{{Weisskopf} {et~al.}(2022){Weisskopf}, {Soffitta}, {Baldini},
  {Ramsey}, {O'Dell}, {Romani}, {Matt}, {Deininger}, {Baumgartner},
  {Bellazzini}, {Costa}, {Kolodziejczak}, {Latronico}, {Marshall}, {Muleri},
  {Bongiorno}, {Tennant}, {Bucciantini}, {Dovciak}, {Marin}, {Marscher},
  {Poutanen}, {Slane}, {Turolla}, {Kalinowski}, {Di Marco}, {Fabiani},
  {Minuti}, {La Monaca}, {Pinchera}, {Rankin}, {Sgro'}, {Trois}, {Xie},
  {Alexander}, {Allen}, {Amici}, {Andersen}, {Antonelli}, {Antoniak},
  {Attin{\`a}}, {Barbanera}, {Bachetti}, {Baggett}, {Bladt}, {Brez}, {Bonino},
  {Boree}, {Borotto}, {Breeding}, {Brienza}, {Bygott}, {Caporale}, {Cardelli},
  {Carpentiero}, {Castellano}, {Castronuovo}, {Cavalli}, {Cavazzuti},
  {Ceccanti}, {Centrone}, {Citraro}, {D'Amico}, {D'Alba}, {Di Gesu}, {Del
  Monte}, {Dietz}, {Di Lalla}, {Persio}, {Dolan}, {Donnarumma}, {Evangelista},
  {Ferrant}, {Ferrazzoli}, {Ferrie}, {Footdale}, {Forsyth}, {Foster},
  {Garelick}, {Gunji}, {Gurnee}, {Head}, {Hibbard}, {Johnson}, {Kelly},
  {Kilaru}, {Lefevre}, {Roy}, {Loffredo}, {Lorenzi}, {Lucchesi}, {Maddox},
  {Magazzu}, {Maldera}, {Manfreda}, {Mangraviti}, {Marengo}, {Marrocchesi},
  {Massaro}, {Mauger}, {McCracken}, {McEachen}, {Mize}, {Mereu}, {Mitchell},
  {Mitsuishi}, {Morbidini}, {Mosti}, {Nasimi}, {Negri}, {Negro}, {Nguyen},
  {Nitschke}, {Nuti}, {Onizuka}, {Oppedisano}, {Orsini}, {Osborne}, {Pacheco},
  {Paggi}, {Painter}, {Pavelitz}, {Pentz}, {Piazzolla}, {Perri},
  {Pesce-Rollins}, {Peterson}, {Pilia}, {Profeti}, {Puccetti}, {Ranganathan},
  {Ratheesh}, {Reedy}, {Root}, {Rubini}, {Ruswick}, {Sanchez}, {Sarra},
  {Santoli}, {Scalise}, {Sciortino}, {Schroeder}, {Seek}, {Sosdian}, {Spandre},
  {Speegle}, {Tamagawa}, {Tardiola}, {Tobia}, {Thomas}, {Valerie}, {Vimercati},
  {Walden}, {Weddendorf}, {Wedmore}, {Welch}, {Zanetti}, \&
  {Zanetti}}]{Weisskopf2022}
{Weisskopf}, M.~C., {Soffitta}, P., {Baldini}, L., {et~al.} 2022, Journal of
  Astronomical Telescopes, Instruments, and Systems, 8, 026002

\end{thebibliography}

\end{document}